\documentclass[]{spie}  

\usepackage{amsmath,amsfonts,amssymb}
\usepackage{graphicx}

\title{Deep Residual Dense U-Net for Resolution Enhancement in Accelerated MRI Acquisition}

\author[a]{Pak Lun Kevin Ding}
\author[b]{Zhiqiang Li}
\author[c]{Yuxiang Zhou}
\author[a]{Baoxin Li}
\affil[a]{School of Computing, Informatics, and Decision Systems Engineering \newline
    Arizona State University, Tempe, AZ 85281}
\affil[b]{Dept. of Neuroradiology, Barrow Neurological Institute, Phoenix, AZ 85013}
\affil[c]{Dept. of Radiology, Mayo Clinic Arizona, Phoenix, AZ 85054}

\authorinfo{
Further author information:\\
P.L.K Ding: kevinding@asu.edu\\
Z. Li: Zhiqiang.Li@BarrowNeuro.org\\
Y. Zhou: Zhou.Yuxiang@mayo.edu\\
B. Li: baoxin.li@asu.edu\\
Accepted to SPIE Medical Imaging, February 2019
}

\begin{document} 
\maketitle

\begin{abstract}
Typical Magnetic Resonance Imaging (MRI) scan may take 20 to 60 minutes. Reducing MRI scan time is beneficial for both patient experience and cost considerations. Accelerated MRI scan may be achieved by acquiring less amount of k-space data (down-sampling in the k-space). However, this leads to lower resolution and aliasing artifacts for the reconstructed images. There are many existing approaches for attempting to reconstruct high-quality images from down-sampled k-space data, with varying complexity and performance. In recent years, deep-learning approaches have been proposed for this task, and promising results have been reported. Still, the problem remains challenging especially because of the high fidelity requirement in most medical applications employing reconstructed MRI images. In this work, we propose a deep-learning approach, aiming at reconstructing high-quality images from accelerated MRI acquisition. Specifically, we use Convolutional Neural Network (CNN) to learn the differences between the aliased images and the original images, employing a U-Net-like architecture. Further, a micro-architecture termed Residual Dense Block (RDB) is introduced for learning a better feature representation than the plain U-Net. Considering the peculiarity of the down-sampled k-space data, we introduce a new term to the loss function in learning, which effectively employs the given k-space data during training to provide additional regularization on the update of the network weights. To evaluate the proposed approach, we compare it with other state-of-the-art methods. In both visual inspection and evaluation using standard metrics, the proposed approach is able to deliver improved performance, demonstrating its potential for providing an effective solution.
\end{abstract}


\section{Introduction}
Magnetic resonance imaging (MRI) is among the most important imaging methods for medical diagnosis. Obtaining fully sampled MRI data requires relatively long scan time. To shorten MRI scan time for improved patient experience and reduced cost, researchers have investigated accelerated MRI acquisition. One basic idea for acceleration is to under-sample in k-space, which may cause aliasing in the reconstructed images. Parallel MRI \cite{sense,grappa} and compressed sensing (CS) MRI \cite{cs,csmri} are two popular techniques. A representative technique of the parallel MRI, generalized auto-calibrating partial parallel acquisition (GRAPPA) \cite{grappa}, uses interpolation to fill the missing k-space data with the surrounding data from all the coils, while CS-MRI randomly samples the k-space data for approximating the original image.

Approaches using low-rank matrix completion technique to solve the CS-MRI/parallel MRI problem were also investigated. Representatives include SAKE \cite{sake} and the annihilating filter based low-rank Hankel matrix approach (ALOHA) \cite{aloha}. However, these algorithms have high complexity and k-space data are required during the reconstruction, making them impossible for cases with only image domain inputs.

In recent years, deep learning has become one of the most important tools for visual computing research \cite{fasterrcnn,maskrcnn}, with great performance in image classification \cite{alexnet}, segmentation \cite{unet}, recognition \cite{resnet}, super resolution \cite{sr}, etc. Therefore, some researcher started to utilize deep-learning techniques for medical image reconstruction. Wang et al \cite{accelmri} trained a convolutional neural network (CNN) to learn the mapping from the aliased image to the original fully sampled reconstruction. The output of the network can be used as an initial guess or regularization term in conventional CS approaches. In ref.\citenum{parmri}, the authors proposed a multilayer perceptron for parallel MRI, and in ref.\citenum{varmri} the researchers applied CNN on CS algorithm. Kang et al \cite{ct} applied the CNN technique on computed tomography (CT), etc.

The authors of a recent paper \cite{drmri} used U-Net \cite{unet} with residual learning to learn the relationship between the aliased and original images, and the proposed framework outperforms traditional methods like SENSE and ALOHA. However, since U-Net is originally developed for medical image segmentation, it is likely that directly using it on image reconstruction may not give us the best performance. For this reason, we propose Residual Dense U-Net (RD-U-Net), a U-Net based deep neural network to further improve the quality of the reconstructed image. Inspired by DenseNet \cite{densenet}, a residual dense block for refinement is introduced to improve the quality of the reconstructed image.
Furthermore, we impose the Fourier constraint into the loss function. Experimental results show that, both visually and numerically, our proposed architecture has a better performance.

\section{Related Works}
\begin{figure} []
   \begin{center}
   \begin{tabular}{c} 
   \includegraphics[width=\linewidth]{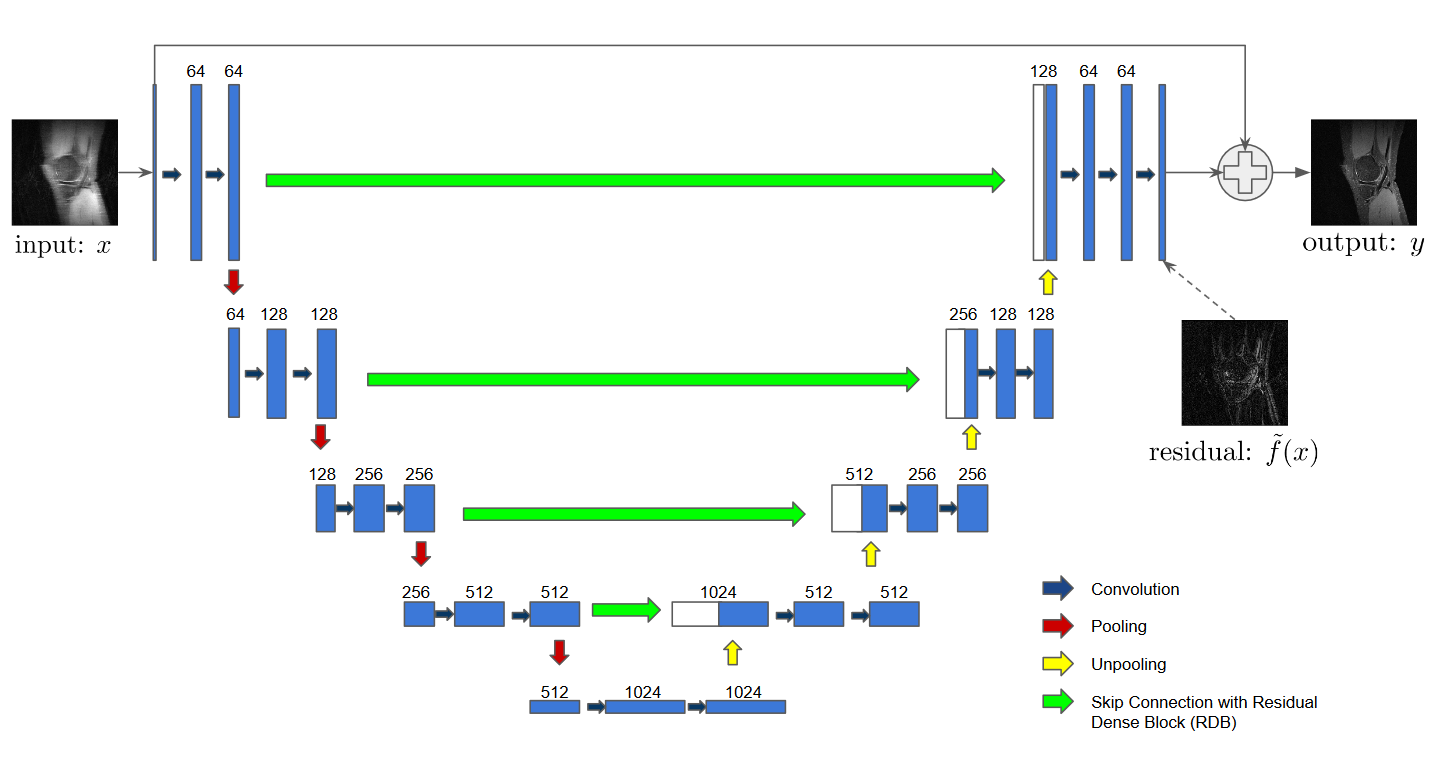}
   \end{tabular}
   \end{center}
   \caption{
        The illustration of our proposed RD-U-Net.
        Blue arrow: 3x3 convolution + Batch Normalization + Nonlinear activation;
        Red arrow: 2x2 max pooling with stride 2;
        Yellow arrow: 2 x 2 up-convolution;
        Green arrow: skip connection with residual dense block.
   }
   \label{fig:unet}
\end{figure}
In this section, we review the works that related to our proposed RD-U-Net model for accelerated MRI reconstruction. We note that there is a line of work on super-resolution using sparse represent ion (e.g., \cite{understandingsparse}), which has delivered superior performance for super-resolution in natural images. However, such work is not amenable for the task of MR image reconstruction in this paper since the degradation involves structured aliasing.

\subsection{U-Net}
U-Net \cite{unet} is first proposed for biomedical image segmentation,
which incorporates skip connection and downsampling/upsampling layers.
These skip connection intend to extract local information,
while the encoding/decoding procedure provide global information.
While obtaining state-of-the-art results,
U-Net is applied to other visual computing field like accelerated MRI reconstruction \cite{drmri} and pansharpening \cite{pan}.

\begin{figure} []
   \begin{center}
   \begin{tabular}{c} 
   \includegraphics[width=0.6\linewidth]{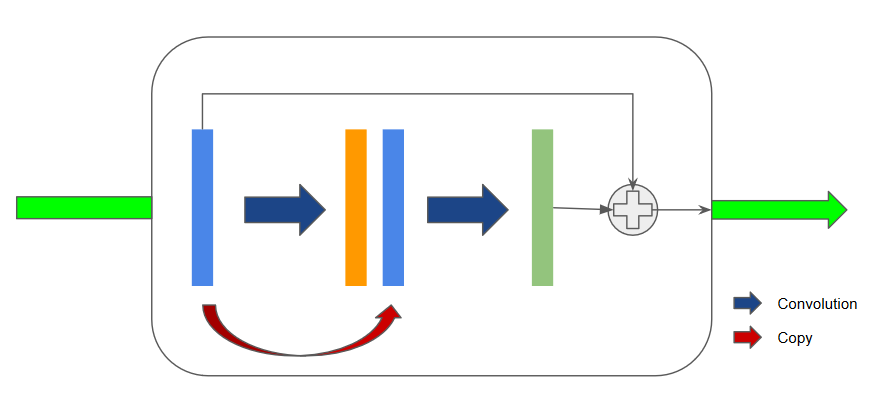}
   \end{tabular}
   \end{center}
   \caption{
        The illustration of the proposed Residual Dense Block (RDB).
        Blue arrow: 3x3 convolution + Batch Normalization + Nonlinear activation;
        Red arrow: Skip Connection;
        "Dense” part:
        The input (blue) is first convolved,
        and concatenate to itself,
        and then the second convolution is applied;
        “Residual” part:
        The input and output of the dense part are summed together to learn the residual. 
   }
   \label{fig:rdb}
\end{figure}
\subsection{Residual learning}
Optimization is an important procedure to deep learning.
A major challenge for the optimization is the gradient vanishing problem.
To overcome this issue,
the concept of residual learning is introduced in the residual network (ResNet) \cite{resnet}.
In this model,
a shortcut connection (skip connection) is used in every basic residual block,
which makes the gradient flow in the networks is relatively stable.
It is also shown that in ref.\citenum{landscape}, 
adding skip connection leads to the flatness of loss surfaces.
ResNet has provided very promising results on many applications.

\subsection{Densely Connected Convolutional Networks}
Many works have shown that the networks perform better if there are connections between layers close to the input and the ones close to the output.
The authors in ref.\citenum{densenet} propose Dense Convolutional Network (DenseNet),
which connects each layer to every other layer by using skip connections.
Differ from the traditional convolutional network,
every layer in DenseNet takes the feature maps from all preceding layers as inputs,
and its output feature maps are served as inputs for the subsequent layers.
DenseNet achieves state-of-the-art performance in a lot of real world problems.


\section{Proposed Approach}
For traditional super resolution,
patch-based \cite{sr,sr1} approaches are usually used.
However,
for accelerated MRI reconstruction,
the aliasing artifacts are of global nature.
To diminish the global artifacts,
we can use the whole image as an input.
The authors of the ref.\citenum{drmri} use U-Net with residual learning to handle this problem.
To be precise,
let $x$ be the input image,
and $\tilde{f}$ be the function represented by the U-Net,
the output of the model can be defined as:
\begin{equation}
    y = f(x) = x + \tilde{f}(x)
\end{equation}
where $f$ is the function representing the whole network.
In this case $\tilde{f}(x) = y - x$,
and since x and y are the low resolution and high resolution image respectively,
$\tilde{f}$ maps $x$ to the residual.
Learning such $\tilde{f}$ leads to faster convergence\cite{drmri}.

The proposed network architecture is illustrated in Fig. \ref{fig:unet}. Similar to the U-Net-based approach \cite{drmri},
it consists of an encoding path and a decoding path. The encoding path is a traditional convolutional neural network, which consists of two 3 x 3 convolutions, each followed by a batch normalization (BN) layer and a nonlinear activation layer. After that, a 2 x 2 max pooling layer with stride 2 is applied for down sampling, and the number of channels is doubled after the down sampling.
For the decoding path, at every stage it consists of a 2 x 2 deconvolution which up sample the feature map and reduce the number of channels by half.  After up-sampling, feature maps from the same level in the encoding path are fed to the Residual Dense Block, and the corresponding output is concatenated, followed by two 3 x 3 convolutional layers, a BN layer and a nonlinear activation layer. A 1 x 1 convolutional layer is used at the final to map all the features to a single channel.

\subsection{Refinement Using Residual Dense Block}

U-Net has its limitations for extracting high frequency data.
Since the high frequency content are only contained in the upper part of the network (the earlier stages of the encoding part, and the later stages of the decoding part),
The network is not deep enough to extract the high frequency features.
Inspired by DenseNet \cite{densenet},
we introduce Residual Dense Block (RDB) to refine the feature map.
RDB is formed by the “dense” part and “residual” part.
For the “dense” part,
the input is first passed through a convolutional layer, a batch normalization layer and a nonlinear activation layer,
where the number of filters used in the convolutional layer is the same as the number of channels of its input.
The output is then concatenated with the input of the RDB,
and passed through another convolutional layer again,
and reduce the number of channels by half.
At the end the input of the RDB is added to the output of the “dense” part,
to form the “residual” part.
Fig. \ref{fig:rdb}(a) is an illustration.

Instead of using the plain skip connection (copying the feature maps from the encoding part to the decoding part), adding a refinement is a more reasonable choice, as theoretically, the plain skip connection is a special case of the RDB (by setting all the weights of the second convolution in RDB to be zero).

\subsection{Employing Fourier Constraints}
\begin{figure} []
   \begin{center}
   \begin{tabular}{c} 
   \includegraphics[width=\linewidth]{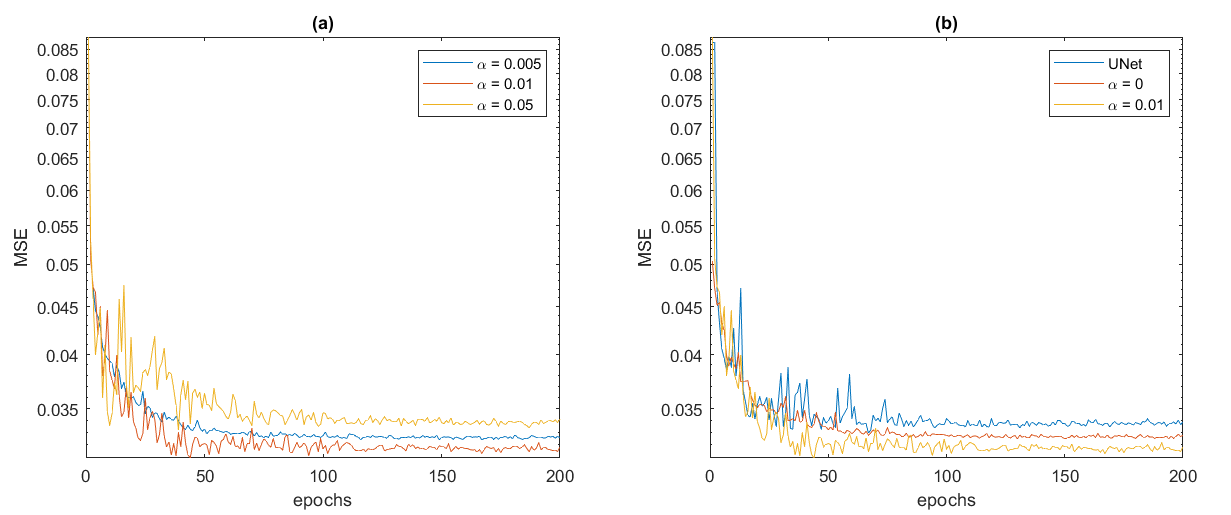}
   \end{tabular}
   \end{center}
   \caption{
        The plots for the testing loss versus number of epochs trained in a trial.
        (a) the testing loss for different value of $\alpha$;
        (b) the testing loss for U-Net,
        RD-U-Net without Fourier constraint (denoted by $\alpha = 0$) and with Fourier constraint (denoted by $\alpha = 0.01$).
   }
   \label{fig:perf}
\end{figure}

\begin{table}[]
\begin{center}
\begin{tabular}{c|cccc}
\hline
    & GRAPPA & U-Net & RD-U-Net($\alpha = 0$) & RD-U-Net($\alpha = 0.01$) \\ \hline \hline
MSE & $0.1483 \pm 0.0052 $ & $0.0338 \pm 0.0003$ & $0.0326 \pm 0.0002$ & $0.0319 \pm 0.0002$ \\ \hline
\end{tabular}
\end{center}
\caption{
The MSE (mean and standard deviation) for different models on 5 trials:
GRAPPA, U-Net, RD-U-Net without Fourier constraint (denoted by $\alpha = 0$), RD-U-Net with Fourier constraint (denoted by $\alpha = 0.01$).
}
\label{tab:result}
\end{table}

$L_2$ loss is usually used in image reconstruction tasks,
which is defined as follows:
\begin{equation}
\label{eqt:loss1}
    \min \|y-f(x)\|_2
\end{equation}
where $f$ is the mapping represented by the neural network.
However, given that the degradation of the images are come from the missing of columns (or rows) of in the K-space data, we can use this prior to improve the performance, by using the following loss:
\begin{equation}
\label{eqt:loss2}
    \min \|y - f(x)\|_2 + \alpha \| F(y) - F(f(x)) \|_1
\end{equation}
where $y$ and $x$ represent the ground truth and the degraded image, respectively, $f$ is the mapping represented by the neural network, $F$ is the inverse Fourier transform, and $\alpha$ is a constant.
As we expect the lost information only comes from part of the rows in the Fourier domain,
we use $L_1$ norm for the Fourier regularization term.
With this loss function we can effectively regularize the network learning by the error coming from those missing k-space data.

\section{Experiments}
\label{sec:exp}

In this section,
we introduce the dataset we use,
followed by the network settings, and comparative studies.
\subsection{Magnetic Resonance Dataset}
We use the fully sampled knee datasets from mridata.org to evaluate our RD-U-Net. There are 20 datasets, and the data were acquired in Cartesian coordinate on a GE clinical 3T scanner, with the following parameters:  Receiver Bandwidth = 50.0, number of coils = 8, acquisition matrix size = 320 $\times$ 320, FOV = 160mm $\times$ 160mm $\times$ 153.6mm, number of slices = 256, TR = 1550ms, TE = 25ms, FA = 90, sequence type = SE. For each dataset, we only use the 150 slices from the central part (the 51st to 200th slices). We randomly pick 1 dataset for training,
and 1 dataset for testing. For data augmentation, we generate 8 times more training samples by rotating and reflecting the images. The original k-space data are retrospectively down sampled by 4 times with 16 ACS (auto-calibration signal, 5\% of total PE).
The sampling pattern, a high resolution reconstructed image and a low resolution reconstructed image (by zero-filled) are shown in Fig. \ref{fig:gt}.

As the images are acquired under different scan conditions,
normalization is necessary for the preprocessing of the data.
We applied the following transformation for every single images:
\begin{equation}
    x \leftarrow \frac{x - mean(x)}{std(x)}
\end{equation}
After the transformation,
the pixels in an image will have zero mean and unit variance.

\subsection{Experimental Protocols and Results}
We train the network for 200 epochs with batch size = 3.
Stochastic gradient descent (SGD) is used with initial learning rate = 0.02 and momentum = 0.5.
For every 20 epochs the learning rate is decreased by half.
We use PoLU \cite{polu} as our activation function.
For calculating the error, we use mean square error (MSE), which can be defined as:
\begin{equation}
    MSE = \frac{\|y - f(x)\|_F^2} {N}
\end{equation}
where $y$ and $x$ represent the ground truth and degraded images,
$N$ represent the number of pixels,
$f$ denotes the function represented by the learned neural network,
$\| \cdot \|_F$ represents the Frobenius Norm.

\begin{figure} []
    \begin{center}
        \begin{tabular}{c}
            (a) \hspace{3.9cm} (b) \hspace{3.9cm} (c)    \\
            \includegraphics[width=0.8\linewidth]{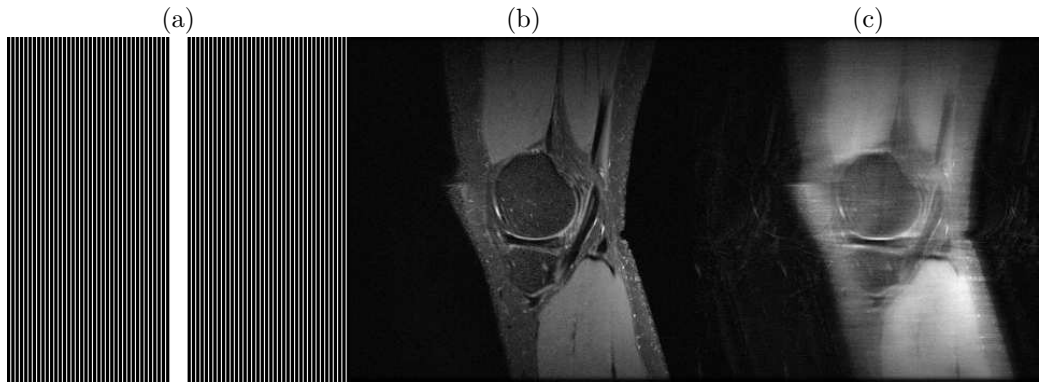}
        \end{tabular}
   \end{center}
   \caption{
        The figures for
        (a) the sampling pattern (4x accelration, with 16 ACS, 5\% of total PE);
        (b) the reconstructed image from fully sampled k-space data;
        (c) the reconstructed image from zero-filled under sampled k-space data.
   }
   \label{fig:gt}
\end{figure}

\begin{figure} []
   \begin{center}
   \begin{tabular}{c}
            (a) \hspace{3.6cm}
            (b) \hspace{3.6cm}
            (c) \hspace{3.6cm} (d)   \\
            \includegraphics[width=\linewidth]{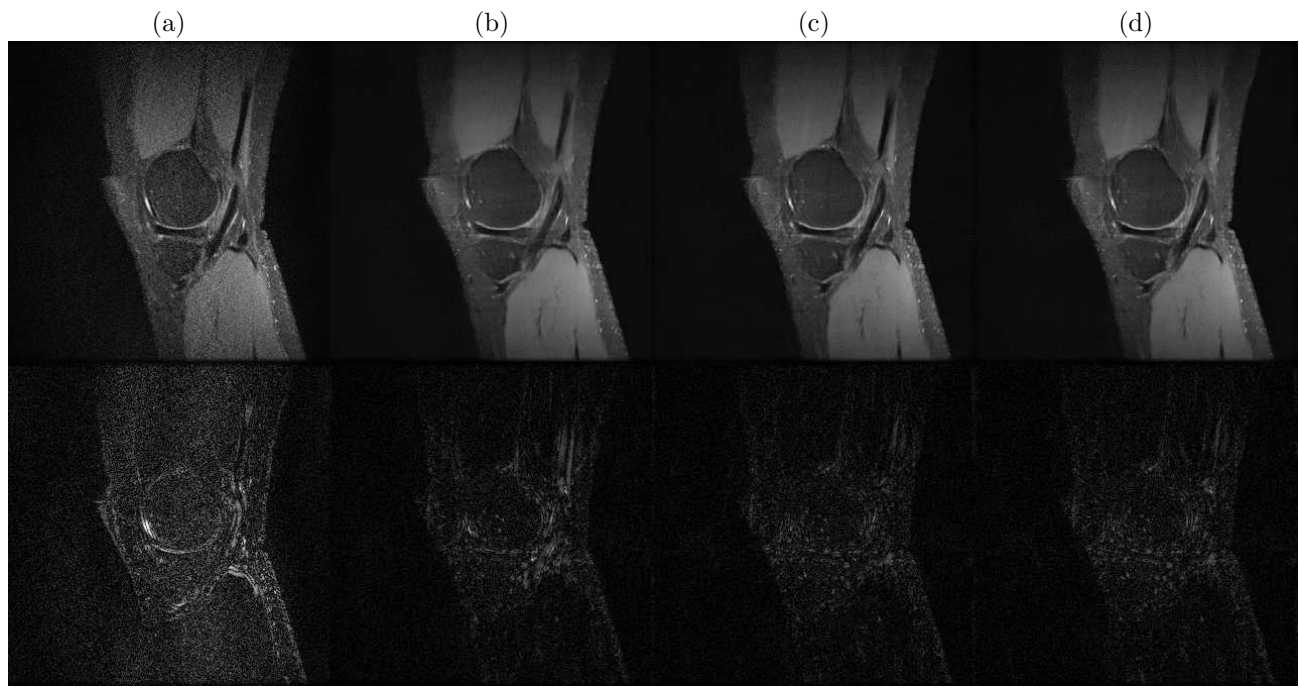}
        \end{tabular}
   \end{center}
   \caption{
        The figures for
        (a) the results from GRAPPA;
        (b) the results from U-Net \cite{drmri};
        (c) the results from RD-U-Net;
        (d) the results from RD-U-Net with Fourier constraint.
        Row 1: the reconstructed images;
        Row 2: the difference between the reconstructed images and the ground truth.
   }
   \label{fig:result}
\end{figure}

To determine the value of $\alpha$,
we first test the our model with $\alpha = 0.05, 0.01, 0.005$ with 1 trial for each value,
using the loss function stated in Eq. \ref{eqt:loss2}.
We use the $\alpha$ with the least MSE for the remaining experiments.
The plot of the testing loss for different $\alpha$ is in Fig. \ref{fig:perf}(a).
To evaluate the performance of the RD-U-Net model,
we compared the reconstruction results with GRAPPA \cite{grappa} and U-Net \cite{drmri}.
Tab. \ref{tab:result} report the results for 5 different runs,
and Fig. \ref{fig:result} illustrates sample visual results.
For the zero-filled reconstruction,
there exists a lot of aliasing artifacts (see Fig. \ref{fig:gt}(c)).
Although GRAPPA removes the aliasing artifacts,
the reconstructed image is still with a lot of noise (Fig.\ref{fig:result}(a)).
The U-Net approach is better than GRAPPA,
but the result is still lack of details (Fig. \ref{fig:result}(b)).
The proposed RD-U-Net provides a better reconstructed image with the presentation of residual dense block.
In Fig.\ref{fig:result}(c),
we can see that the result is sharper at the edges,
and clearer for the details.
The Fourier constraint is also useful for obtaining a better reconstruction,
as shown in Fig. \ref{fig:perf}(b),
adding the such regularization leads to lower MSE.

The network was implemented using Pytorch 0.4.0 with python 3.6.3 on Ubuntu 16.04.
All the experiments were performed on a computer with nVidia GTX 1080 GPU and Intel Xeon E5-2603 CPU, although the code was not optimzed to with respect to the particular computer hardware.

The reconstruction time for GRAPPA is about 20 seconds.
The training time for the U-Net, RD-U-Net(without/with Fourier constraint) are about 13, 15 and 15 hours respectively.
The reconstruction time for all the three compared neural networks are less than 1 second.

\section{CONCLUSION}
In this paper, we proposed a new architecture to approximate the fully sampled MR images from the down sampled MR ones. The architecture is based on U-Net, and it achieves low NMSE during the reconstruction of MR images, due to the participation of the residual dense refinement and the Fourier regularization. The visual results show that our proposed model is able to reduce more aliasing artifacts.

\bibliography{report} 
\bibliographystyle{spiebib} 

\end{document}